\documentclass[runningheads]{svmult}

\usepackage{makeidx}   
\usepackage{graphicx}  
\usepackage{subeqnar}  
\usepackage{multicol}  
\usepackage{physprbb}  
\makeindex             

\def\gtrsim{\mathrel{\hbox{\rlap{\hbox{\lower4pt\hbox{$\sim$}}}\hbox{$>$}}}}
\def\sun{\hbox{$\odot$}}

\begin{document}
\def\farcs{\hbox{$.\!\!^{\prime\prime}$}}
\def\arcsec{\hbox{$^{\prime\prime}$}}
\def\micron{\hbox{$\mu$m}}
\def\uchii{UCH{\sc ii}}
\title*{Infrared Observation of Hot Cores \footnote{Based on observations
        collected at the European Southern
        Observatory, La Silla, Chile, Proposal IDs 62.I-0530,67.C-0359}}
\toctitle{Infrared Observation of Hot Cores}
%
%
\titlerunning{Infrared Observation of Hot Cores}
%
\author{Bringfried Stecklum\inst{1}
\and Bernhard Brandl\inst{2}
\and Markus Feldt\inst{3}
\and Thomas Henning\inst{4}
\and Hendrik Linz\inst{1}
\and Ilaria Pascucci\inst{4}
}
\authorrunning{Bringfried Stecklum et al.}
%
%
\institute{Th\"uringer Landessternwarte Tautenburg, Sternwarte 5,
                D--07778 Tautenburg
\and Center for Radiophysics \& Space Research, Cornell University, Ithaca, NY 14853
\and Max-Planck-Institut f\"ur Astronomie, K\"onigstuhl 17, D--69117 Heidelberg
\and Astrophysikalisches Institut und Universit\"ats--Sternwarte,
           Friedrich-Schiller-Universit\"at Jena, Schillerg\"a{\ss}chen 2--3,
           D--07745 Jena}

\maketitle              

\begin{abstract}
We report on mid-infrared imaging of hot cores
performed with SpectroCam--10 and TIMMI2. The observations aimed
at the detection of thermal emission presumably associated with the
hot cores. Mid-infrared flux measurements are required to improve the
luminosity and optical depth
estimates for these sources. Results are presented for W3(H$_2$O), G9.62+0.19,
G10.47+0.03, and the possible hot core candidate G232.620+0.996. They illustrate that
the morphology of these sources cannot be described by simple geometries. Therefore,
line-of-sight effects and considerable extinction even at mid-infrared wavelengths must not
be neglected.
\end{abstract}

\section{Introduction}
Hot cores (HCs) are the suspected birthplaces of high-mass stars (M$\gtrsim 8$\,M$_{\sun}$)
\cite{ppiv00}. Based on mm/submm interferometric results which provide sufficient
angular resolution to separate HCs from neighbouring ultracompact H{\sc ii}
regions (\uchii s), first attempts were made to derive properties of the embedded
high-mass stars as well as the surrounding envelope using infall models in combination with
1D radiative codes \cite{osor99}. However, these results relied on the mm/submm
part of the spectral energy distribution (SED) only. We performed
infrared (IR) observations of HCs to provide flux densities (or at least
upper limits) for the atmospheric 10 and 20\,\micron{} spectral windows in order
to complete the coverage of their SEDs with sub-arcsecond beam sizes and to investigate
their morphology. 

\section{Observations and Data Reduction}
The observations were carried out with SpectroCam--10 (SC\,10) on the 5-m 
Hale\footnote{Observations at the Palomar Observatory
were made as part of a continuing collaborative agreement between the California
Institute of Technology and Cornell University.} and TIMMI2 on the ESO 3.6-m telescopes.
SC\,10 is the Cornell-built 8--13~$\mu$m
spectrograph/camera \cite{hayward93} which utilizes a
Rockwell 128$\times$128 Si:As BIB array, providing a circular field of view
(FOV) in
imaging mode of 16\arcsec{} (pixel size 0\farcs25).
The SC\,10 imaging was performed in December 1998 and June 1999.
The images were filtered using a wavelet algorithm \cite{pantin96}
to enhance the signal-to-noise ratio (SNR).
TIMMI2 is the new ESO thermal imaging multi-mode instrument
covering the wavelength range from 5--20\,\micron{} \cite{reim00}.
Its 320$\times$240 Si:As BIB array manufactured
by Raytheon provides an unprecedented FOV in imaging mode of 
$64\arcsec\times48\arcsec$ (pixel scale
0\farcs2) at 10 and 20\,\micron{} and $96\arcsec\times72\arcsec$
(pixel scale 0\farcs3) at
4.7\,\micron. The observations were carried out in March 2001.

\section{Results}
\subsection{W3(H$_2$O)/W3(OH)}
\vspace*{-0.5cm}
\begin{figure}[h]
\begin{center}
\includegraphics[width=.6\textwidth]{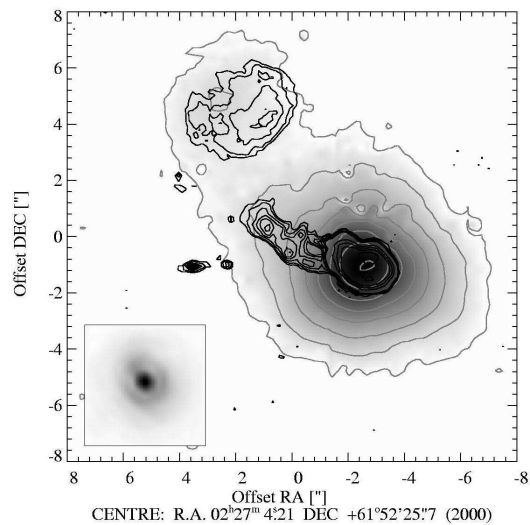}
\caption[]{SC\,10 11.7\,\micron{} image of W3(OH) with contours
of the 11.7\,\micron{} (grey) emission and of the 3.6\,cm radio continuum (black,
from \cite{wilne95}). W3(H$_2$O) is located at the offset position
[+3\farcs6,-1\farcs0]. The inset image of $\alpha$~Tau indicates the
beam size.}
\end{center}
\label{eps1}
\vspace*{-0.5cm}
\end{figure}
The HC W3(H$_2$O) is located $\sim6$\arcsec{} east of the \uchii{}
 W3(OH) at a distance of 2.2\,kpc. The proper motion of its H$_2$O masers
\cite{alco92} and the associated radio continuum jet \cite{reid95} suggest
recent outflow activity. Interferometric radio observations of W3(H$_2$O)
\cite{wyro99} yielded
H$_2$ column densities of up to 1.5$\times10^{24}\,$cm$^{-2}$ and
rotation temperatures in the range of 160--200\,K. The detection of the HC
at mid-infrared (MIR) wavelengths was claimed by \cite{keto92}. 
Fig.~1 
shows the SC\,10 image of 
W3(H$_2$O)/W3(OH) with black contours delineating the 3.6\,cm radio continuum
\cite{wilne95}. W3(H$_2$O) is at the offset position
[+3\farcs6,-1\farcs0] and traced by its radio continuum jet, but is not detected
at 11.7\,\micron{}. The feature identified as W3(H$_2$O) by \cite{keto92} is
presumably the weak cometary \uchii{} northeast of W3(OH). The radio position of
the \uchii{} W3(OH) served as as astrometric reference. The comparison of the
spatial extent of W3(OH) with the standard star
$\alpha$~Tau (inset of 
Fig.~1)
shows that it is clearly resolved.\\
The temperature and size derived from molecular line
interferometry for W3(H$_2$O) suggest that it should be a bright object in
the MIR, with a predicted 11.7\,\micron{} flux density of
$\sim2000$\,Jy. Our failure to detect
it despite a much better sensitivity (3$\sigma$ point source detection limit of
6\,mJy) than \cite{keto92} implies a large
amount of cold dust in front of the hot core. A detailed analysis of the SEDs of
W3(H$_2$O)/W3(OH) is given in \cite{stec01}.

\begin{figure}[t]
\begin{center}
\includegraphics[width=.6\textwidth]{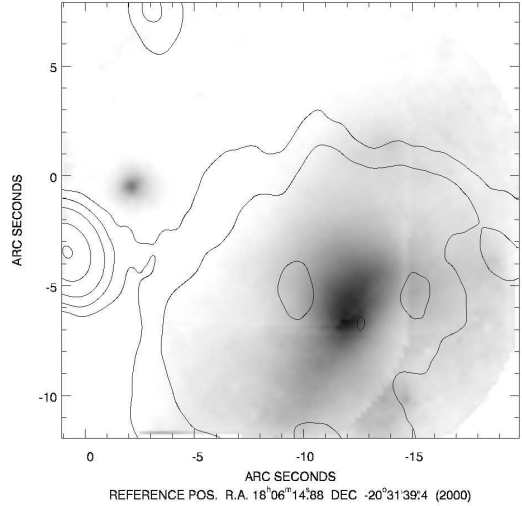}
\caption[]{SC\,10 11.7\micron{} image of G9.62+0.19 with contours of
the 1.3\,cm radio continuum (from \cite{cesa94}).}
\end{center}
\label{eps2}
\vspace*{-0.5cm}
\end{figure}
\subsection{G9.62+0.19}
The object G9.62+0.19 (IRAS18032-2032) comprises several \uchii s at a distance of 5.7\,kpc.
Recently, weak radio continuum emission has been detected \cite{test00} at the location of
the HC (component F) as well as outflow activity \cite{hofn01}. Remarkably, this object seems to
be associated with 2.2\,\micron{} emission \cite{test98}, a fact which can
be hardly reconciled with spherical models of HCs. Such models imply
extinction values which should prevent the detection
of the HC at near-infrared (NIR) wavelengths. 
Fig.~2 
shows the 11.7\,\micron{} SC\,10
image of G9.62+0.19 with contours of the 1.3\,cm radio continuum
\cite{cesa94}. The bulk of the 11.7\,\micron{} emission stems from the
extended radio component B which was used as astrometric reference. Surprisingly,
no thermal emission was seen from the \uchii{} D which again implies a considerable
optical depth even at MIR wavelengths for this source. This result concerning
component D  contradicts that of \cite{buiz00} due to differing astrometry. A
pointlike source is obvious close to the location of the HC
(offset position [-2\farcs0,-1\farcs0]). This emission is almost coincident with the
2.2\,\micron{} feature. The poster contribution of Linz et al. (this volume)
discusses this source and the relation of the observed IR emission to
the recently detected molecular outflow \cite{hofn01} in more detail. Possibly,
IR radiation arising from the HC escapes in the outflow lobe inclined
towards the observer where the optical depth is lower.


\begin{figure}[t]
\hspace*{2cm}
\includegraphics[width=.65\textwidth]{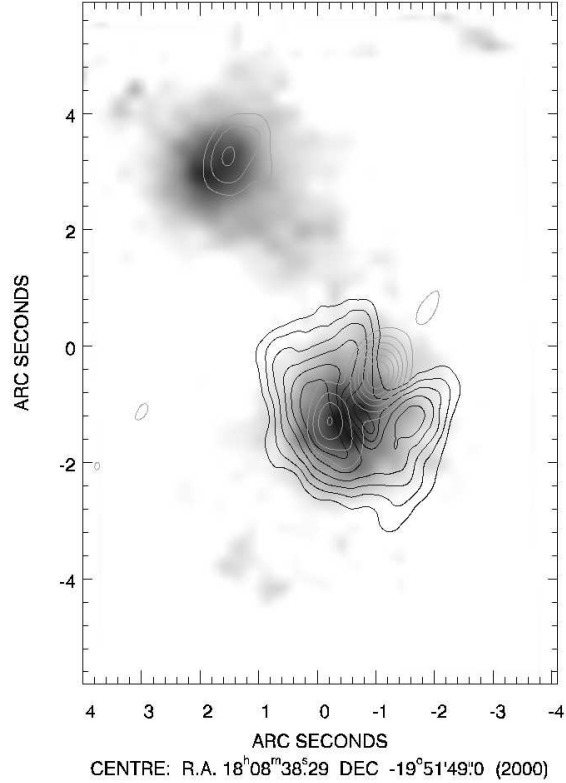}
\caption[]{SC\,10 11.7\micron{} image of G10.47+0.03 with contours of
the 6\,cm radio continuum (grey, from \cite{wc89}) and of the NH$_3(4,4)$
line (black, from \cite{cesa98}).}
\label{eps3}
\vspace*{-0.5cm}
\end{figure}
\subsection{G10.47+0.03}
According to the radial brightness temperature profile based on interferometric
maps of the NH$_3$(4,4) line, the HC G10.47+0.03 (IRAS18056-1952)
seemed to be one of the rare cases with {\em established} evidence for internal
heating by embedded OB stars contrary to external heating by adjacent \uchii s
\cite{cesa98}. 
Two stars from the corresponding 2MASS images of the region were also detected
with SC\,10 at 11.7\,\micron{}. They provide astrometric reference with sub-arcsecond
precision. MIR emission from two sites in the G10.47+0.03 region was detected. These
sources are not seen
on the  2.2\,\micron{} 2MASS image (10$\sigma$ detection limit of 1.2\,mJy). 
Fig.~\ref{eps3} displays the SC\,10 11.7\,\micron{} image with contours of
the 6\,cm radio continuum (grey, from \cite{wc89}) and of the NH$_3(4,4)$
emission (black, from \cite{cesa98}). The HC is located at the offset
position [0\farcs0,--1\farcs6]. To the northwest of the HC, the NH$_3$ emission is diminished
due to absorption by the \uchii{}s in the foreground. The centroid of the 11.7\,\micron{}
radiation almost coincides with the center of the NH$_3$ emission. A
possible configuration which explains this morphology might be a cometary
\uchii, similar to G29.96--0.02 and its adjacent HC (which are seen side-on),
with the line of sight slightly inclined with respect to the symmetry axes of
the H{\sc ii} region. The fact that there is only one MIR peak associated with
the HC suggests that the multiple structure in the radio continuum (3
components are present in
the map of \cite{cesa98}) is presumably due to locally enhanced plasma density
and not caused by a few widely distributed high-mass
stars. The lack of NH$_3$ emission for the second MIR source (offset position
[+1\farcs5,+3\farcs0]) indicates that this is presumably a more evolved \uchii{}.


\begin{figure}[h]
\hspace*{2cm}
\includegraphics[width=.6\textwidth]{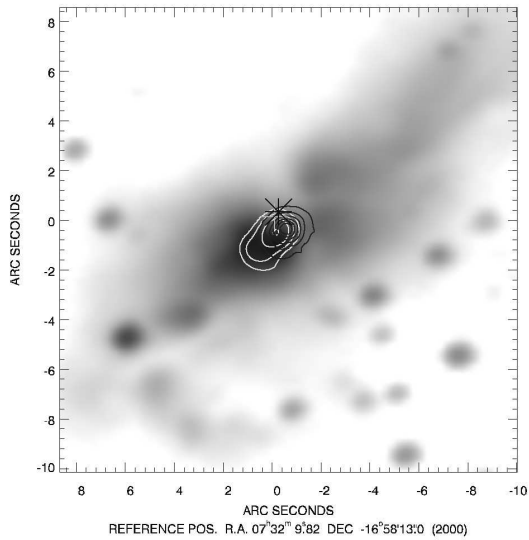}
\caption[]{2.2\micron{} SOFI image of G232.620+0.996 with contours of
the 4.7\,\micron{} (white) and the 11.9\,\micron{} (black) emission. The
reference position is defined by the location of the masers. The black
asterisk
marks the location of the illuminating source derived from our 2.2\,\micron{}
polarization map.}
\label{eps4}
\vspace*{-0.5cm}
\end{figure}
\subsection{G232.620+0.996}
The \uchii{} G232.620+0.996 (IRAS07299-1651) is
associated with OH  and CH$_3$OH  masers \cite{casw98},\cite{wals98}
which are signs of newly formed massive stars. The \uchii{} is 1\farcs5
northwest of the maser position \cite{wals98}. These objects are located
at the southern border of a dense core seen in the 1.3\,mm map taken with
the SEST \cite{klein01}. Our 2.2\,\micron{}
polarimetric  map obtained with SOFI at the ESO-NTT revealed that the bipolar-like extended
emission (see Fig.~4) at the southern rim of the dense core is dominated by scattered light.
The location of the illuminating
source was derived by minimizing the sum of the scalar products between
polarisation and radius vector as a function of source position. It is marked
in Fig.~4 by the black asterisk, and situated very close to the masers and the \uchii.
The large FOV of TIMMI2 covered the target and  HD\,60068 (situated 47\arcsec{} northwest of it)
simultaneously which allowed to establish accurate astrometry.
A resolved IR source was found close to the maser position. 
Fig.~4 also displays contours of the 4.7\,\micron{} and 11.9\,\micron{} image taken with
TIMMI2. It can be noticed that the emission peaks of both wavelengths are spatially
offset, indicating a strong gradient in the optical depth in
the immediate neighbourhood of the illuminating source. 

\section{Acknowledgements}
This work was supported by DFG grants STE~605/17-1 and STE~605/18-1.

%

\end{document}